\documentstyle[preprint,aps]{revtex}

\begin{document}
\draft
\title{\bf   Trapped surfaces in cosmological spacetimes.}
\author{Edward Malec$^{+*}$ and Niall \'O Murchadha$^{+}$}
\address{$^+$Erwin Schr\"odinger Institute, Vienna, Austria}
\address{$^+$on leave from the Physics Department, University College, Cork,
Ireland}
\address{$^*$on leave from  Institute of Physics,  Jagellonian University,
30-059  Cracow, Reymonta 4, Poland}

\maketitle
\date{\today}
\begin{abstract} We investigate the formation of trapped surfaces
in cosmological spacetimes, using constant mean curvature slicing.
Quantitative criteria for the formation of trapped surfaces
demonstrate that cosmological  regions enclosed by trapped surfaces may have
matter density  exceeding significantly the background matter density
of the flat and homogeneous cosmological model.
\end{abstract}
\section{INTRODUCTION}

In our previous work (\cite{1}) we investigated the formation of trapped
surfaces in various cosmological models.  Recently we have found  a
particularly useful formulation (\cite{2}) of the spherically symmetric
Einstein
constraint equations that allowed us to improve our early estimates (\cite{3})
for conditions determining the appearance of trapped surfaces.
In the present paper we apply the new formalism to spherically symmetric
cosmologies. As a result we find  stronger  criteria in  spacetimes  that were
investigated previously and, more importantly, we  are able to  deal with
hyperbolic universes where our previous attempts have failed.

The order of the article is as follows. The first  section presents the
formalism. In  Section 2 we  deal with the main results. Section 3 shows that
regions enclosed by trapped surfaces must be invisible to external observers.
The last Section contains conclusions of which the most important is that
energy density inside cosmological regions enclosed by trapped surfaces
may exceed significantly the average energy  density of the flat and
homogeneous cosmological model.

There exist three homogeneous spherically symmetric cosmologies. The three are

i) the closed (k=1) cosmology with metric
 \begin{equation}
ds^2 = -  d\tau^2 + a^2(\tau )[dr^2 +  \sin^2 r  d\Omega^2],
\label{1}
\end{equation}

ii) the open flat (k=0) cosmology with metric
 \begin{equation}
ds^2 = -  d\tau^2 + a^2(\tau )[dr^2 +  r^2  d\Omega^2],
\label{2}
\end{equation}

iii)
 \begin{equation}
ds^2 = -\  d\tau^2 + a^2(\tau )[dr^2 +  \sinh^2 r  d\Omega^2],
\label{3}
\end{equation}
where $d\Omega^2 = d\theta^2 + \sin^2\theta d\phi^2$ is the standard line
element on the unit sphere with with the angle variables $0\le \phi < 2\pi $
and $0\le \theta \le \pi $.

The geometric part of the initial data set  of the Einstein equations
consists of the intrinsic three-geometry
and the extrinsic curvature $K_{ab}$ which is essentially the first time
derivative of the metric, all given at some  time  (say $\tau =0$). The
intrinsic geometries are respectively
 \begin{equation}
  a^2(\tau )[dr^2 +  \sin^2 r  d\Omega^2],
\label{4}
\end{equation}

 \begin{equation}
  a^2(\tau )[dr^2 +  r^2  d\Omega^2],
\label{5}
\end{equation}

  \begin{equation}
  a^2(\tau )[dr^2 +  \sinh^2 r  d\Omega^2],
\label{6}
\end{equation}
and in each case the extrinsic curvature is pure trace

  \begin{equation}
K_{ab}=Hg_{ab}
\label{7}
\end{equation}
where $H$ is a time dependent function that is constant on each slice
$\tau =const $.  It is called the Hubble constant and it is  given by
$H={\partial_{\tau } a\over a}$.

In the general case  initial data consist of the quartet $(g_{ij}, K_{ij},
\rho, J_i)$ where $g_{ij}$ is the intrinsic metric, $K_{ij}$ is the extrinsic
curvature, $\rho $ is the matter energy density and $J_i$ is the matter
current density. These cannot be given arbitrarily but must satisfy the
constraints
 \begin{equation}
^{(3)}{\cal R}-K_{ij}K^{ij}+(trK)^2=16\pi \rho
 \label{8}
\end{equation}
 \begin{equation}
 \nabla_iK^{ij}-\nabla^jtrK =-8\pi J^j
 \label{9}
\end{equation}
where $^{(3)}{\cal R}$ is the scalar curvature of the intrinsic metric.

The momentum constraint, (\ref{9}), is identically satisfied in the case of
homogeneous cosmologies (with $J_i=0$) and the hamiltonian constraint,
(\ref{9}), reduces to
\begin{equation}
  16 \pi \rho  ={6k\over  a^2} +6H^2
\label{10}
\end{equation}
where $k$ is $1, 0, -1$ in the closed, flat and hyperbolic cosmologies,
respectively. Thus we
can conclude that all slices of the constant coordinate time have a
uniform energy density $\rho_0$ which is at rest.

In this article we wish to consider data for spherically symmetric cosmologies
which   either in the large approximate the standard cosmologies or
asymptotically approach them. In all cases we will make the assumption
that the initial slice is chosen so that the trace of the extrinsic
curvature is constant on the slice. In order to retain the link with
homogeneous cosmologies we define $tr K= 3H $.

The initial data we consider is a spherically symetric set  consisting
of a   three-metric
 \begin{equation}
ds^2_{(3)} =   a^2 dr^2 + b^2(r)  f^2(r)  d\Omega^2,
\label{11}
\end{equation}
an   extrinsic curvature
 \begin{equation}
K_r^r=H+K(r),~~~ K_{\theta }^{\theta }=H-K/2,~~~ K_{\phi }^{\phi }=H-K/2
\label{12}
\end{equation}
an energy density $\rho (r)$ and a  current density $j_i$. The function
$f(r)$ will be one of the set
$\sin (r), r, \sinh(r)$, depending on the type of cosmology.

There are some useful geometric quantities that can be defined. One of them
is the proper distance from the center of symmetry given by $dl= adr$.
The  Schwarzschild (areal) radius $R$ is given by $R=bf$.
  The  mean curvature of a centered two-sphere as embedded in
an initial three dimensional hypersuface  is
\begin{equation}
p= {2\partial_lR\over   R}.
\label{13}
\end{equation}

In a general spacetime  we may investigate the geometry by considering
the propagation of various beams  of lightrays through a space-time.
These beams in general will shear and either expand or contract; a number
of (optical) functions will be required to describe their propagation.
In a spherically symmetric spacetime   we
focus our attention to light rays moving orthogonally to two-spheres
centered around a center of symmetry. We need only two functions. These are
the divergence of future directed light rays
\begin{equation}
\theta ={2\over R}|_{out}{d\over d\tau }_{out}R
\label{14a}
\end{equation}
and the divergence of past directed light rays
\begin{equation}
\theta ' ={-2\over R}{d\over d\tau }_{in }R
\label{14b}
\end{equation}
where  ${d\over d\tau }_{out }$ is the derivative along future-pointing
outgoing radial null rays and ${d\over d\tau }_{in }$ is   the derivative
along  future-pointing ingoing radial null rays. One interesting property of
$\theta $ and $\theta'$ is that they can be expressed purely in terms of
initial data on a spacelike slice. In the spherically symmetric case we
have
\begin{equation}
\theta = p-K_r^r+trK=p-  K +2H,
\label{15a}
\end{equation}
and
\begin{equation}
\theta '  =p+K_r^r-trK=p+  K -2H.
\label{15b}
\end{equation}
This means that $\theta $ and $\theta '$ are three-dimensional scalars.
They are not four-scalars, since they depend on a choice of affine
parameters along the null rays.

In the   homogeneous universes
 we find that $ pR =2$, $pR= 2\cos (r)$ and $pR= 2cosh (r)$
in the $k=0, 1,-1$ cases respectively and
\begin{equation}
R\theta   = 2+2RH = 2 + 2\sqrt{{8\pi \rho_0 \over 3}}R,\ \ R\theta '= 2-2RH
\label{16a}
\end{equation}
for k=0,
\begin{eqnarray}
R\theta   = &&2\cos (r)+2RH=2\cos (r)+2aH\sin (r)
=2\cos (r)+2\sqrt{({8\pi \rho_0a^2 \over 3}-1)}\sin (r) ,\nonumber\\
&& R\theta '= 2\cos (r)-2\sqrt{({8\pi \rho_0a^2 \over 3}-1)}\sin (r)
\label{16b}
\end{eqnarray}
for k=1,
\begin{eqnarray}
R\theta   = &&2\cosh (r)+2RH=
2\cosh (r)+2\sqrt{({8\pi \rho_0a^2 \over 3}+1)}\sinh (r) , \nonumber\\
&&R\theta '= 2\cosh (r)-2\sqrt{({8\pi \rho_0a^2 \over 3}+1)}\sinh (r)
\label{16c}
\end{eqnarray}
for k=-1.

A surface on which $\theta $ is negative is called, after Penrose {\cite{4}),
a future trapped surface
and a surface on which $\theta'$ is negative is called a past trapped
surface. The occurrence of such surfaces in a spacetime  is an indication of
the fact that the gravitational collapse is well advanced. In the case of
homogeneous closed cosmologies future trapped surfaces
exist for any $ r> \cot^{-1}(aH)$.  In neither k=0 nor k=-1 is $R\theta $
ever negative if $H>0$.

 In this article we  consider  a universe that  is
homogeneous in the large but that it is
dotted  with  numerous spherical inhomogeneities, far from each the metric
approaches the background  metric of a homogeneous universe. If we center
our coordinate system  at a particular lump we expect that   optical
scalars approach the values given in (\ref{16a}, \ref{16b}, \ref{16c}) far
away from the lump.
In the case of closed cosmologies this limiting value is expected to
be met for values of the coordinate radius $r$ much less than $\pi /2$.

We assume local flatness at the origin, i. e.,
$\lim_{ R\rightarrow 0}R\theta =\lim_{ R\rightarrow 0}R\theta'=2$ although
this condition can be relaxed to allow for a conical singularity there,
i. e., $0<\lim_{ R\rightarrow 0}R\theta ,\lim_{ R\rightarrow 0}R\theta'\le 2$.

\section{  MAIN CALCULATIONS.}

 The spherical initial data must satisfy the constraints, which read, in
 terms of functions $\theta $ and $\theta'$
\begin{eqnarray}
\partial_l ( \theta R)=&&-8\pi R(\rho -  j )
-{1\over 4R}[2(\theta R)^2-\theta R\theta' R-4-12\theta RHR]
\label{17}
\end{eqnarray}
\begin{eqnarray}
\partial_l ( \theta' R)=&&-8\pi R(\rho +  j )
-{1\over 4R}[2(\theta' R)^2-\theta R\theta' R-4+12\theta RHR]
\label{18}
\end{eqnarray}
where $j=j_l$ is   the  radial component of the matter current density
normalized so that $j^2=j^kj_k$. We can manipulate equations (\ref{17})
and  (\ref{18}) to obtain
\begin{eqnarray}
\partial_l ( \theta' R\theta R)=&&-8\pi \Bigl( \rho (\theta' R+\theta R)+
j(\theta R - \theta 'R)\Bigr)
-{1\over 2R}[( \theta R\theta' R-4)(\theta' R+\theta R)].
\label{19}
\end{eqnarray}

Let us now  assume that the total matter satisfy the dominant energy
condition, i. e.,  $ \rho \ge |j| $.  Assume that $\theta R\theta 'R>4$
at a particular point. Consider first the situtation  where both $\theta R$
and $\theta' R$  are positive. Then  $(\theta' R+\theta R)>
(-\theta' R+\theta R)$ and $\rho (\theta' R+\theta R)+j(\theta' R+\theta R)
\ge 0$.  This means that  both terms of (\ref{19}) are nonpositive
  and the derivative of the product $\theta R\theta 'R$ is  negative. On the
  other hand, when both  $\theta R$ and $\theta 'R$ are negative and their
product is greater than 4, then
$\rho (\theta' R+\theta R)+j(\theta' R+\theta R) <0$ and the first term in
(\ref{19}) is positive. The second term becomes also positive, so that
$\partial_l(\theta R\theta 'R)>0$. Thus in both cases if $\theta R\theta 'R>4$
then  $\partial_l(\theta R\theta 'R)\ne 0$.

Let us now consider the expressions for the product of the two scalars
$\theta R\theta 'R$ in each of the three homogeneous cosmologies.
We get
\begin{equation}
R\theta R\theta '  = 4-4R^2H^2
\label{20a}
\end{equation}
for k=0,
\begin{equation}
R\theta R\theta '  = 4\cos^2 (r)-4R^2H^2
 \label{20b}
\end{equation}
for k=1 and
\begin{equation}
R\theta R\theta '  = 4\cosh^2 (r) -4({8\pi \rho_0a^2 \over 3}+1)\sinh^2 (r)=
4([1  -{8\pi \rho_0a^2 \over 3}]\sinh^2 (r))
\label{20c}
\end{equation}
for k=$-1$.

In each of these cases we have $R\theta R\theta '=4$ at the origin and
never more  than  4. We are considering initial geometries that  locally
are flat and asymptotically approach the homogeneous cosmologies, so
that both at the origin and  far from the center the product $R\theta
R\theta '$ does not  exceed 4.  If it were to achieve a maximal value
greater  than 4 somewhere in between, then its derivative would have
to vanish; but that is excluded in the preceding analysis. Therefore we
have proven:

{\bf Lemma 1.} Assume that matter satisfies the dominant energy condition and
 that spherical cosmological data   are locally flat at the center
 and are asymptotic to
 any of standard homogeneous cosmological models. Then
 $$ R\theta R\theta ' \le 4.$$
Remarks:

i) The above statement is true for any regular slice, with arbitrary
(i. e., nonconstant on a part of a slice) tr$K$, assuming that the
slice is asymptotic to a homogeneous constant mean curvature slice.

ii) It implies the positivity of the Hawking mass  on a sphere centered
around a symmetry center; $2M_H=R(1- {R\theta R\theta ' \over 4})$
cannot become negative on a fixed slice.

Lemma 1 holds true for all three cosmological models.

The main issue that we will address in this paper is the question of the
formation of trapped surfaces due to concentration of matter. The result
will be obtained through a careful analysis of (\ref{17}).  What we do is
multiply  (\ref{17}) by $R$, use (\ref{13}) and write  the resulting
equation in the following form
\begin{equation}
\partial_l ( \theta R^2)=-8\pi R^2(\rho -  j ) +1
+{1\over 2} \theta R\theta' R-{1\over 4} (\theta R)^2 + 3\theta RHR.
\label{22}
\end{equation}
The substitution of (\ref{15a}) and (\ref{15b}) into (\ref{22}) gives
\begin{equation}
\partial_l ( \theta R^2)=-8\pi R^2(\rho - {3H^2\over 8\pi }- j ) +1
+{1\over 4} (p R+KR)^2-R^2K^2+2R^2Hp
\label{23}
\end{equation}
or
\begin{eqnarray}
\partial_l ( \theta R^2)=&&-8\pi R^2(\rho -\rho_0 +{3k\over 8\pi a^2} - j )
+2-(1- {1\over 4} (p R+KR)^2)-R^2K^2+4RH\partial_lR,
\label{24}
\end{eqnarray}

where we used the relation (\ref{10}) to eliminate the $H^2$ term and use the
definition of mean curvature $p$.

Let us integrate  (\ref{24}) from the origin out to a surface $S$.
 We identify
\begin{equation}
 \Delta M = 4\pi \int_0^{L(S)} R^2(\rho -\rho_0 )dl =
 \int_{V(S)}dV(\rho -\rho_0)
\label{25}
\end{equation}
as the excess matter inside a volume $V(S)$ bounded by $S$  and
\begin{equation}
 P= 4\pi \int_0^{L(S)} R^2 jdl = \int_{V(S)}dV j
\label{26}
\end{equation}
as the total radial momentum inside $S$. In this notation, the
aforementioned  integration  yields

\begin{equation}
  \theta R^2|_S= -2(\Delta M-P ) -{3k\over 4\pi a^2}V+2L+{HA\over 2\pi }
 -\int_{V(S)}dV\Bigl( 1 - {1\over 4} (p R+KR)^2+R^2K^2\Bigr)
\label{27}
\end{equation}
where $A$ is the area of the surface $S$ and $L$ is the geodesic distance
of $S$ from the centre.  Below we will prove, in a series of lemmas,
that under some conditions  we can control the sign of the last integral.

{\bf Lemma 2.} Assume k=0, 1 cosmologies which are locally flat.
If the   energy condition  $\rho -\rho_0 -|j|\ge
-{3k\over 4\pi a^2}$
is satisfied out to an asymptotic region then

$$2\ge  | p R+KR|,~~~   2\ge   |p R-KR| .$$

{\bf Lemma 3.}
In a data set that approaches the $k=-1$ locally flat cosmology, if the energy
condition
$\rho -\rho_0 -|j|\ge  {3\over 4\pi a^2}$ is satisfied inside a sphere $S$ then

$$2 >   (p R+KR),~~~2> (pR-KR). $$

Before proving the two lemmas, let us formulate two main results that
give sufficient conditions for  the formation of trapped surfaces.

{\bf  Theorem 1.}
Given data which approaches either the k=0 or the k=1 locally flat  cosmology,
if the   energy condition
$\rho -\rho_0 -|j|\ge
{3k\over 4\pi a^2}$
is satisfied out to an asymptotic region
and if
\begin{equation}
   \Delta M-P \ge  -{3k\over 8\pi a^2}V+L+{HA\over 4\pi }
\label{28}
\end{equation}
at a surface $S$ then $S$ is future  trapped.

{\bf Proof of theorem 1}: the result follows directly from eq. (\ref{27}) and
the estimate of Lemma 2.

{\bf  Theorem 2.}
Assume that normally  ingoing light light rays are everywhere convergent
inside  a volume $V$ bounded by a surface $S$, $\theta '>0$.
Given data which approaches the  $k=-1$ locally flat  cosmology, if the
energy
condition
$\rho -\rho_0 -|j|\ge {3\over 4\pi a^2}$ is satisfied  inside the volume $V$
and
if
\begin{equation}
   \Delta M-P \ge  {3\over 8\pi a^2}V+L+{HA\over 4\pi }
\label{29}
\end{equation}
at the surface $S$ then there exists a surface inside $S$ that is future
trapped.

{\bf Proof of Theorem 2}.  Assume that there is no future  trapped surface
inside
$S$, i. e., $\theta =p+2H-K>0$. Since we also assume that there is no  past
trapped surface, we may conclude that inside $S$ $p-K>-2H, p+K>2H$. We know
that $p$ is positive inside $S$ because we have that $p = (\theta +
\theta')/2$ and each of $\theta$ and $\theta'$ is positive. We also have $2 >
pR - KR > -2HR$ and $2 > pR + KR > 2HR$; the last inequalities  follow from
lemma 3. If $H > 0$ we have that  $pR + KR$ is positive and thus $(p+K)^2R^2
\le
4$  and the last integral of (\ref{27}) is strictly negative. On the other
hand,
if
$H < 0$ we must have that $pR - KR$ is positive and $(pR - KR)^2 \le 4$ but we
could have that $pR + KR$ be negative.  This can only happen while $K$ is
negative since we know that $p$ is positive. In this case we write the
integrand of (\ref{27} as $1 -{1 \over 4}(pR - KR)^2 - pKR^2 + K^2R^2$.
This is clearly nonnegative. Thus  we also have in this case that the
last term in (\ref{27}) is negative. This contradicts the assumption that there
is no  trapped surface. Hence, under the assumptions of   Theorem 2, there must
exist a trapped surface inside
$S$.

In order to prove lemmas 2 and 3 we shall return to equations (\ref{17})
and (\ref{18}) and write them in terms of $Rp, RK$ and $RH$. (\ref{17})
can written as
\begin{equation}
\partial_l (  p R-KR)=-8\pi R(\rho -  j -{3H^2\over 8\pi })
-{1\over 2R}(Rp-RK)^2+{1\over 4R}(Rp-RK)(Rp+RK) +{1\over R}
\label{30}
\end{equation}
and (\ref{18})  as
\begin{equation}
\partial_l (  p R+KR)=-8\pi R(\rho +  j -{3H^2\over 8\pi })
-{1\over 2R}(Rp+RK)^2+{1\over 4R}(Rp-RK)(Rp+RK) +{1\over R}.
\label{31}
\end{equation}
We will prove first the upper bound of $pR+KR, pR-KR$, simultaneously
for both Lemma 2 and 3; this part of the proof does not depend on
the type of a cosmological spacetime. Also, as it will become clear,
the energy condition shall be imposed only inside a sphere $S$ if we  are
interested in finding the estimate inside $S$
(as opposed to the estimations from below that require the  global
assumption  made in Lemma 2). According to the conditions
made in lemmas, the first term of either equation (\ref{30}) or (\ref{31})
is  nonpositive. We show that in the situation of interest the
remainders of each of the  equations are also nonpositive.

At the  center of symmetry  the quantities $pR+KR, pR-KR$
are equal to 2, for all types of cosmology. This means that right hand
sides of either  (\ref{30}) or (\ref{31}) must be nonpositive and that the
quantities in question  start from the origin with the value 2 and start
to decrease as soon as they meet either positive  $\rho +j -{3H^2\over 8\pi }$
or $ \rho -  j -{3H^2\over 8\pi }$.

Let us assume that further out one of the two, say $pR+KR$, rises up to
2 with  $pR-KR$ lagging behind. In this case  we can write the non-material
part of the right hand side of  (\ref{31}) as follows
\begin{equation}
-{1\over 2}(Rp+RK)^2+{1\over 4}(Rp-RK)(Rp+RK) +1=
-1+{1\over 2}(Rp-RK)\le 0.
\label{32}
\end{equation}
Because the material part of (\ref{31}) is nonpositive, we get that
$\partial_l(Rp+RK) \le 0$ so that $pR+KR$ cannot exceed 2. A similar argument
can be made for $pR-KR$. Thus Lemma 3 and the upper  bound of Lemma 2  are
proven; as is clear from the above derivation, in order to have a bound
that is valid   inside a sphere $S$ we need the energy condition that is
imposed  only inside $S$.

The same reasoning can be applied to complete the proof of Lemma 2.
We will show, that if one of the two quantities in question reaches
the value -2, then at least one of them must be less than -2, thus
breaking either the demand of geometries being asymptotic to a homogeneous
cosmology in the sense expressed in  equations (\ref{16a})
and (\ref{16b}).

In order to show this we need the global energy condition
of Lemma 2. Let us assume  that there exists a point where  $pR+KR= -2$,
with $pR-KR\ge pR+KR$. Then the nonmaterial part of Eq. (\ref{31})
reads
\begin{equation}
-{1\over 2}[2(Rp+RK)^2+{1\over 4}(Rp-RK)(Rp+RK) +1\le 0
\label{33}
\end{equation}
Eq. (\ref{31}) implies now (assuming the energy condition) that  $pR+KR$
has to become more  negative, if $pR+KR<pR-KR$ and
may stay at -2 in the case of equality only if the matter contribution
exactly cancels. However, if we can impose an outer boundary condition such
that  $pR+KR\ge -2$ then we get a contradiction.  A similar argument works
for $pR-KR$. The outer boundary condition is guaranteed in the cases of
interest. Cosmological spacetime dotted with
inhomogeneities have the property that asymptotically $pR+KR$   and  $pR-KR$
approach values given by  (\ref{16a}) and  (\ref{16b}) which must be
strictly bigger than -2.
  That ends the proof of Lemma 2.

   It is interesting that we obtain an exact  criterion with   the
constant 1; this suggests that the above theorem  constitute a  part of a
more complex true statement that can be formulated for general nonspherical
spacetimes. It suggests also that $M(S)$ is a sensible measure of the energy
of a gravitational  system that  might appear as a part of a
quasilocal energy measure in nonspherical systems.

It is clear that the analysis performed here can include cases where the
sources
are distributions rather than classical functions; in particular, we have no
difficulty with shells of matter. All we get on crossing the shell is a
downward step in $\theta$ and $\theta'$. More interestingly, we can extend
the analysis to include conical singularities at the origin (\cite{5}),
in a way analogous  to that described in (\cite{2}).

\section{ CONFINING PROPERTY OF TRAPPED SURFACES.}

In this Section we show that a region enclosed by trapped surfaces
cannot be seen by external observers. This fact has been proven (without
referring to the Cosmic Censor Hypothesis) by Israel (\cite{6}) . Here we will
present a different version of the proof that is based on a 1+3 decomposition
of a spacetime (as opposed to the proof of Israel, who used a 2+2
decomposition).

We need the evolution part of the Einstein equations
and the lapse equation. These are
 \begin{equation}
 \partial_t(\delta K_r^r -2H(t))= {3\alpha \over 4}(\delta K_r^r)^2
 -{\alpha p^2\over 4} -{p\over \sqrt{a}}\partial_r\alpha +{\alpha \over R^2}
 +8\pi \alpha T_r^r +3\alpha H^2 -3H\delta K_r^r
 \label{34}
\end{equation}
and
 \begin{equation}
\nabla_i \partial^i  \alpha =\alpha \Bigl( {3  \over 2}(\delta K_r^r)^2
4\pi  (\rho + T_i^i) +3  H^2  \Bigr) +3\partial_tH.
 \label{35}
\end{equation}
In addition we need the evolution equation of the mean curvature
$p$ of centred spheres
 \begin{equation}
 \partial_t  p={\partial_r\alpha \over \sqrt{a}}(-\delta K_r^r +2H)+
8\pi \alpha {j_r\over \sqrt{a}} +{p\alpha \over 2}(\delta K_r^r-2H)
 \label{36}
\end{equation}

Using these equations we can find the full time derivative of $\theta $
along a trajectory of null geodesics normal to centred spheres
 \begin{equation}
(\partial_t +{\alpha \over \sqrt{a}})\theta =
{\partial_r\alpha \over \sqrt{a}} \theta ' \theta -
8\pi \alpha (-2{j_r\over \sqrt{a}}+\rho +T_r^r) -\alpha  \theta ^2+
3\alpha H\theta .
 \label{37}
\end{equation}
Take now an apparent horizon, i. e., a  centred sphere $S$  of vanishing
$\theta (S)$; (\ref{37})
implies that photons that start from $S$ will forever remain inside an
apparent horizon,  if the  strong energy condition $-2{j_r\over \sqrt{a}}+
\rho +T_r^r\ge 0$ is assumed.  Hence apparent horizons move faster than light
in cosmological spacetimes (in contrast with asymptotically flat spacetimes,
where they can eventually stabilize to the speed of light); they act
as one-way membranes for non-tachyonic matter. This means that outside
observers cannot detect any information from any inside region that is
enclosed by a trapped surface. The only way to draw any conclusions
about a piece of a spacetime that is enclosed by a trapped surface is
through the observation of "long-wave" effects - through the attractive
force that large massive objects exert on their surrounding.

\section{DISCUSSION.}

Cosmological trapped surfaces that we discuss  in preceding sections
can, if they exist, accumulate an enormous amount of energy.
Typically, as we have shown, the matter content of a trapped surface having
a geodesic radius $L$ is of the order $L$ plus the background energy
$M_H={3H^2V\over 8\pi }$ (we neglect here the effects related to the
possibility of a nonzero curvature of the space-like slice and the
surface term
${HA\over  2\pi }$). Assume that there exists a trapped surface with a proper
radius of the order of 1000 megaparsecs.  Then its excess energy is of the
order
of  1000 (in units of megaparsecs). The present value of the Hubble constant is
about $50{km\over s*megaparsec}$ or (in units in which the speed of light c=1)
${1\over 6000 megaparsec}$. Therefore the expected value of the background
energy inside the above ball is of the order $0.5*({1\over 6000})^2
megaparsec =14 magaparsec$,
which is about $10^2$  times less than the energy content that is
needed in order to form, say, a spherical massive shell that creates
a trapped surface.  We include this crude calculation just to point out that
the
formalism of general relativity does allow for cosmological regions with high
concentrations of matter that are in principle invisible by external
observers.
\acknowledgements

This work was partially supported by Forbairt grant SC/94/225 and the KBN grant
2 PO3B 090 08.


\begin{references}
\bibitem{1} U. Brauer and E. Malec, Phys. Rev. {\bf D45}, R1836(1992);
E. Malec and N. \'O Murchadha, Phys. Rev.  {\bf D47}, 1454(1993);
E. Malec, U. Brauer and N. \'O Murchadha, Phys. Rev.  {\bf D49}, xxxx(1994).
\bibitem{2} E. Malec and N. \'O Murchadha,  Phys. Rev. {\bf D50}, R6033(1994).
\bibitem{3} P. Bizo\'n, E. Malec and N. \'O Murchadha, Phys. Rev. Lett.
{\bf 61} 1147(1988); Class. Quantum Grav. {\bf 6}, 961 (1989);
{\bf 7}, 1953(1990).
\bibitem{4} R. Penrose,  Phys. Rev. Lett. {\bf 14}, 57 (1965).
\bibitem{6} W. Israel, Phys. Rev. Lett. {\bf 56}, 86(1986).
 \bibitem{5} J. Guven and N. \'O Murchadha, to be published.
\end{references}
\end{document}